\documentclass{aa}  

\usepackage{graphicx}
\usepackage{textcomp}
\usepackage{txfonts}
\usepackage{cuted}
\usepackage{answers}
\usepackage{natbib}
\begin{document} 

   \title{Distribution of spin-axes longitudes and shape elongations of main-belt asteroids}

   \author{H. Cibulkov\'a \inst{1} \and J. \v{D}urech \inst{1} \and D. Vokrouhlick\'y \inst{1} \and M. Kaasalainen \inst{2}
          \and
          D. A. Oszkiewicz \inst{3,4}
          }
       
   \institute{Institute of Astronomy, Faculty of Mathematics and Physics, Charles University,
              V Hole\v{s}ovi\v{c}k\'ach 2, 180 00 Prague 8\\
              \email{cibulkova@sirrah.troja.mff.cuni.cz}
         \and Department of Mathematics, Tampere University of Technology, PO Box 553, 33101 Tampere, Finland
         \and Lowell Observatory, 1400 W Mars Hill Rd, Flagstaff, 86001 AZ, USA
	 \and Astronomical Observatory Institute, Faculty of Physics, Adam Mickiewicz University, S\l{}oneczna 36, Pozna\'n 60-286, Poland
             }

   \date{Received ***, 2016; Accepted 5 October 2016}

 
  \abstract
   {Large all-sky surveys provide us with a lot of photometric data that are sparse in time (typically few measurements per night) and can be potentially used for the determination of shapes and rotational states of asteroids. The method which is generally used to derive these parameters is the lightcurve inversion. However, for most asteroids their sparse data are not accurate enough to derive a unique model and the lightcurve inversion method is thus not very efficient.}
    {To fully utilize photometry sparse in time, we developed a new simplified model and applied it on the data from the Lowell photometric database. Our aim was to derive spin axis orientations and shape elongations of asteroids and to reconstruct distributions of these parameters for selected subpopulations to find if there are some differences.}
    {We model asteroids as geometrically scattering triaxial ellipsoids. Observed values of mean brightness and the dispersion of brightness are compared with computed values obtained from the parameters of the model -- ecliptical longitude $\lambda$ and latitude $\beta$ of the pole and the ratios $a/b$, $b/c$ of axes of the ellipsoid. These parameters are optimized to get the best agreement with the observation.}
    {We found that the distribution of $\lambda$ for main-belt asteroids is not uniform (in agreement with findings of Bowell et al., \citeyear{2014M&PS...49...95B}, M\&PS, 49, 95) and is dependent on the inclination of orbit. Surprisingly, the non-uniformity of $\lambda$ distribution is larger for asteroids residing on low-inclination orbits. We also studied distributions of $a/b$ for several groups of asteroids and found that small asteroids ($D<25\,$km) are on average more elongated than large ones.}
    {}

   \maketitle

\section{Introduction}

The amount of photometric data of asteroids has been growing rapidly during the last years. These data are a source of information about shapes and rotational states of asteroids. Currently, the main method for determination of spin states and shapes of asteroids from photometry is the inversion of lightcurves, that was developed by \citet{2001Icar..153...24K} and \citet{2001Icar..153...37K}. Models obtained with this method are stored in the Database of Asteroid Models from Inversion Techniques (DAMIT, \v{D}urech et al. \citeyear{2010A&A...513A..46D}), which now contains models for 907 asteroids. The photometric data can be formally divided into two groups: (i) data dense in time, which well sample the rotational period and which are typically used for the lightcurve inversion method, and (ii) data sparse in time (few measurements per night), which are produced by all-sky surveys, such as Pan-STARRS, Catalina or LONEOS. \citet{2004A&A...422L..39K}, \citet{2005EM&P...97..179D} and \citet{2007IAUS..236..191D} showed that it is possible to get the solution of the inverse problem from sparse photometry if the data are of good quality (noise $\lesssim 5\%$). New asteroid models were derived also by a combination of dense and sparse photometry (\v{D}urech et al. \citeyear{2009A&A...493..291D}, Hanu\v{s} et al. \citeyear{2011A&A...530A.134H}, Hanu\v{s} et al. \citeyear{2013A&A...551A..67H}, Hanu\v{s} et al. \citeyear{2016A&A...586A.108H}).

The first statistical study of pole orientation of asteroids (based on 20 bodies) was made by \citet{1986Icar...68....1M} who revealed the lack of poles close to the ecliptic plane. That was later confirmed by analysis of \citet{2002aste.conf..113P}, \citet{2002Icar..160...24S} and \citet{2007Icar..192..223K} for slightly less than 100 asteroids. \citet{2011A&A...530A.134H} (using a sample of 206 main belt asteroids) found the dependence of the distribution of ecliptical latitudes $\beta$ on the diameter $D$: for $D\gtrsim60\,$km they found basically isotropic distribution of $\beta$ value with only a slight excess of prograde rotators, while the distribution of $\beta$ value for $D\lesssim 30$~km asteroids was found to have a strong preference to either small or large values indicating pole orientation near the pole of the ecliptic. The lack of poles near the ecliptic is most probably due to the Yarkovsky-O'Keefe-Radzievskii-Paddack (YORP) effect, which can alter the direction of spin axis of asteroids smaller than $\approx40\,$km on a time scale shorter than their collisional lifetime (e.g., Pravec \& Harris \citeyear{2000Icar..148...12P}, Rubincam \citeyear{2000Icar..148....2R}). The distribution of ecliptical longitudes $\lambda$ of spin axes was, however, supposed to be rather uniform. For instance,  \citet{1989aste.conf..805D} came to this conclusion from the simulations of the collisional evolution of asteroids. With the growing number of asteroids for which pole orientation have been determined, a reliable statistics could be achieved and this hypothesis could be tested. However, even for a sample of 206 asteroids \citet{2011A&A...530A.134H} did not reveal any non-uniformity in distribution of $\lambda$, but in the same time the data were too few to indicate meaningful non-uniformities. On the contrary, \citet{2002Natur.419...49S} and \citet{2003Icar..162..285S} revealed a non-uniform pole distribution for $20-35$~km size members
      in the Koronis family. In particular, the prograde-rotating
      asteroids all had ecliptic longitude between $24^\circ$ and
      $73^\circ$. This conundrum was resolved by \citet{2003Natur.425..147V}, who showed that these objects underwent a $2-3$~Gyr
      long dynamical evolution during which the
      YORP effect tilted
      their spin axis near the ecliptic pole. Since YORP also continued
      to decrease the rotation frequency in their model, the spin state
      was captured in the Cassini resonance between the pole
      precession due to the solar torque and the orbit precession due
      to Jupiter-Saturn perturbations. The stationary point of this
      particular secular, spin-orbit resonance is currently at $\simeq
      35^\circ$ ecliptic longitude. Thus all bodies whose spin axes
      librate about this point must have $\lambda$ near this value. More recently, \citet{2014M&PS...49...95B} estimated the ecliptical longitudes $\lambda$ for more than $350\,000$ asteroids of the main belt using the magnitude method (Magnusson \citeyear{1986Icar...68....1M}), based on the variation of brightness with the ecliptical longitude: the maximum of brightness corresponds with the spin axis pointing either toward or opposite from the Earth. Surprisingly, the resulting distribution is clearly non-uniform with an excess of asteroids with $\lambda$ from $30^\circ$ to $110^\circ$ and with minimum for $120^\circ$ to $160^\circ$. 

The success of getting a unique solution of the inverse problem with currently available sparse photometric data (which are not accurate enough) is low, nevertheless, using the distributed computing project Asteroids@home (\v{D}urech et al. \citeyear{2015A&C....13...80D}), which significantly reduces the computational time of the period search, \citet{2016A&A...587A..48D} derived 328 new models from the analysis of Lowell photometric data. This is an impressive, but still small increase in number to allow population-wide study. For this reason, here we describe a new method for determination of the orientations of spin axes and shapes of asteroids to utilize photometric data sparse in time. The uncertainties of spin vectors are large for individual bodies, therefore we work with groups of asteroids and construct distributions of tested parameters, because working with large samples of bodies should smear uncertainties of individual solutions and, if uncorrelated, the results should hold in a statistical sense.

The structure of this paper is following: in Section~2, we describe our model and test its reliability on synthetic data, in Section~3, we apply the model to the photometric data from the Lowell Observatory database and construct the distributions of ecliptical longitudes for main-belt asteroids and for several groups of asteroids. Section~4 deals with distributions of the ratio $a/b$ of axes of asteroids and in Section~5 we summarize the main results.


\section{Model}

In the lightcurve inversion method, all parameters describing the rotational state (the rotational period and the orientation of the spin axis), the shape and the light scattering on the surface are fitted and the unique sidereal rotational period $P$ has to be determined. In the case of dense photometric data, we can substantially reduce the computational time necessary for the determination of $P$ by searching only the interval around the value estimated from dense lightcurves. For sparse data, we usually do not have any estimate of $P$ and we have to search the interval of all possible values, which is time-consuming. Moreover, for majority of asteroids we currently do not have sparse data accurate enough to derive a unique rotational period. Therefore, to fully utilize sparse photometry, we developed a new model, which does not allow to determine the rotational period, but it provides an approximate solution for the orientation of the spin axis and the shape parameters of the asteroid.

We model asteroids as geometrically scattering triaxial ellipsoids ($a\geq b \geq c=1$) rotating about the shortest axis of the inertia tensor. The parameters of the model are ecliptic longitude $\lambda$ and latitude $\beta$ of the pole, and the ratios of axes $a/b$ and $b/c$ of the ellipsoid, alternatively axes $a$ and $b$. The advantage of this model is that the brightness $L$ (which is proportional to the projected area of the illuminated and visible part of the surface) can be computed analytically (Connelly \& Ostro \citeyear{1984STIA...8534424C}):
\begin{equation}
 L\propto\frac{\pi abc}{2}\left(\sqrt{{\bf e}^T M {\bf e}} + \frac{{\bf e}^T M {\bf s}}{\sqrt{{\bf s}^T M {\bf s}}}\right)\,,
 \label{ostro}
\end{equation}
where ${\bf e}$, ${\bf s}$ are unit vectors defining the position of the Earth and the Sun in the asteroid coordinate system of principal axes of the inertia tensor, and
\begin{equation}
M =
\begin{pmatrix}
1/a^2 & 0 & 0\\
0 & 1/b^2 & 0 \\
0 & 0 & 1/c^2
\end{pmatrix}
.
\end{equation}

In a special case of opposition ${\bf e}={\bf s}$, the equation (\ref{ostro}) simplifies to
\begin{equation}
 L\propto \pi abc\sqrt{{\bf e}^T M {\bf e}}\,.
\end{equation}
The direction towards Earth can be described by the rotational angle $\phi$ and aspect angle $\theta$ (i.e. angle between ${\bf e}$ and the direction of the spin axis): 
\begin{equation}
 {\bf e}=[\sin\theta\cos\phi,\sin\theta\sin\phi,\cos\theta]^T\,.
\end{equation}
Having set $c=1$, the squared brightness $L^2$ normalized by the maximal possible value $\pi ab$ is
\begin{equation}
L^2 = \frac{\sin^2\theta \cos^2\phi}{a^2} + \frac{\sin^2\theta \sin^2\phi}{b^2} + \cos^2\theta\,.
\label{abrightness}
\end{equation}
The mean quadratic brightness over one rotational period is then
\begin{equation}
 \langle L^2\rangle = \frac{1}{2\pi} \int_0^{2\pi} L^2 {\rm d}\phi = 1+\frac{1}{2}{\rm sin}^2\theta\left(\frac{1}{a^2}+\frac{1}{b^2}-2\right)\,,
 \label{brightness}
\end{equation}
and the normalized dispersion of squared brightness is 
\begin{eqnarray}
 \eta&=&\frac{\sqrt{{\rm var}(L^2)}}{\langle L^2\rangle}=\frac{\sqrt{\langle (L^2 - \langle L^2\rangle)^2\rangle}}{\langle L^2\rangle} =\nonumber\\
 &=&\frac{a^2-b^2}{\sqrt{8}}\left[\frac{a^2b^2}{\sin^2\theta}+\frac{1}{2}(a^2+b^2-2a^2b^2)\right]^{-1}\,.
 \label{dispersion}
\end{eqnarray}
We used equations (\ref{brightness}) and (\ref{dispersion}), to compute $\langle L^2_{\rm model}\rangle$ and $\eta_{\rm model}$, respectively, for each asteroid and for each of its apparition (we defined apparitions as sets of observations with the gap between these sets at least 100 days).

\begin{figure}
\centering
\includegraphics[width=8cm]{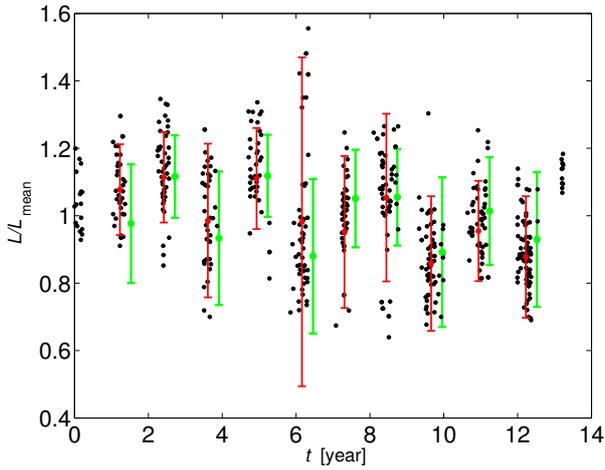}
\caption{Photometric data of the asteroids (511) Davida corrected by the influence of the solar phase angle (black points). Red points with vertical lines denote the observed mean brightness and its dispersion in individual apparitions, green colour denotes the same but calculated quantities for the best-fit model. To normalize $L$, we divided each value by mean value of $L$ calculated over all apparitions.}
\label{oppositions}
\end{figure}

For the observational data, we used the following procedure:
\begin{enumerate}
\item
{We have to remove the dependence on solar phase angle. The changes in brightness in the lightcurve of an asteroid are not only due to the rotation but also due to the geometry of observation. In the model, we assume the case of opposition, that means the solar phase angle $\alpha=0$. For the observational data, we fitted the dependence of the brightness on the solar phase angle $\alpha$ by a linear--exponential dependence similarly as \citet{2011A&A...530A.134H}
\begin{equation}
 g\left(h \exp^{-\alpha/d}-k\alpha+1\right)\frac{1+\cos\alpha}{2}\,,
 \label{phaseangle}
\end{equation}
where $g$, $h$, $d$, $k$ are parameters fitted for each asteroid, and we divided the observed brightness by that function. As an example, the corrected data for asteroid (511) Davida are shown in Fig. \ref{oppositions}.}
\item {Then, we required that there were enough data for each asteroid: at least 20 points in one apparition; and at least five apparitions for one asteroid (on Fig. \ref{oppositions} there are data from 10 apparitions which can be used).}
\end{enumerate}

Computed theoretical values of the mean brightness $\langle L^2\rangle$ and of the dispersion of the brightness $\eta$ can be compared with observation by computation of 
\begin{eqnarray}
\chi^2&=&\chi^2_{\eta}+w\chi^2_{L^2}=\sum_{\rm apparitions}\frac{(\eta_{\rm model}-\eta_{\rm obs})^2}{\sigma_{\eta}^2}+\nonumber\\
&+&w\sum_{\rm apparitions}\frac{(\langle L^2_{\rm model}\rangle/\overline{\langle L^2_{\rm model}\rangle}-\langle L^2_{\rm obs}\rangle/\overline{\langle L^2_{\rm obs}\rangle})^2}{\sigma_{L^2}^2}\,,
\label{chi2}
\end{eqnarray}
where $\sigma$ denotes the standard deviation and $w$ denotes the weight for $\chi^2$ of brightness. To normalize values of calculated and observed mean quadratic brightness, we divided them by $\overline{\langle L^2_{\rm model}\rangle}$ and $\overline{\langle L^2_{\rm obs}\rangle}$, respectively, which are mean values calculated over all apparitions. The value of $w$ is not set in advance and has to be found by testing on known data. Since $\langle L^2\rangle$ and $\eta$ are not gaussian random variables, the $\chi^2$ in relation (\ref{chi2}) is not $\chi^2$-distributed. Nevertheless, we use this $\chi^2$ formalism to define the best solution, which has the minimum~$\chi^2$. 

In passing we note that combining equations (\ref{brightness}) and (\ref{dispersion}) we obtain, for a given asteroid, relation between $\langle L^2\rangle$ and $\eta$:
\begin{equation}
\eta = \frac{1}{\sqrt{2}}\frac{a^2-b^2}{a^2+b^2-2a^2b^2}\left[1-\frac{1}{\langle L^2\rangle}\right]\,.
\end{equation}
This implies that for larger $\langle L^2\rangle$ the model predicts smaller dispersion $\eta$. This is in accord with the intuition that larger brightness corresponds to the pole-on geometry of view (i.e., smaller aspect angle $\theta$). 

To find a model with the best agreement (the lowest $\chi^2$) between the calculated values and the observation, we computed model values on a grid in parameter space: the ecliptical longitude of the pole from $0^\circ$ to $360^\circ$; the latitude from $0^\circ$ to $90^\circ$, both with the step $5^\circ$, and the axes $a$ and $b$, from 1.1 to 4 and from 1 to $a$, respectively, both with step 0.1 (elongation larger than $4:1$ would be unrealistic).
As mentioned above, we corrected observed brightness to the solar phase angle $\alpha=0$, however, the geometry remained unchanged and the aspect angle $\theta$ (which appears in equations (\ref{brightness}) and (\ref{dispersion})) was calculated for each apparition as a mean value:
\begin{equation}
\cos\theta_{\rm mean}=\bf{v}\cdot \bf{e_{\rm mean}}\,,
\label{theta}
\end{equation}
where $\bf{v}=[\cos\beta\cos\lambda, \cos\beta\sin\lambda, \sin\beta]^{\rm T}$ is the vector defining the direction of the spin axis and $\bf{e_{\rm mean}}$ is the mean vector defining the position of the Earth during one apparition. From the relation~(\ref{theta}) we can see that we obtain the same aspect angle for $\lambda$, $\beta$ and $\lambda \pm 180^\circ$ ,$-\beta$, which is the reason why we test $\beta$ only in the interval from $0^\circ$ to $90^\circ$. Relation~(\ref{theta}) also indicates that, for most asteroids, there will be only slightly worse second minimum of $\chi^2$ for $\lambda\pm180^\circ$. For zero inclination of orbit ($e_z=0$), the aspect angle would be the same for $\lambda$ and $\lambda\pm180^\circ$. Due to this ambiguity in $\lambda$, we constructed distributions of $\lambda$ only in the interval $0^\circ$--$180^\circ$ and for $\lambda>180^\circ$ we used modulo $180^\circ$.

\subsection{Testing of the model on synthetic data}

To test our model and confirm its reliability, we created synthetic data: we computed brightness of asteroids using the models from DAMIT database and the Hapke's scattering model (Hapke \citeyear{1981JGR....86.4571H}; Hapke \citeyear{1993tres.book.....H}) with randomly chosen parameters and we assigned this new (synthetic) values to asteroids contained in the Lowell database (to the time of observation and the appropriate geometry). The distribution of poles for this synthetic data was isotropic. 

We added the gaussian noise (we tested noise $\sigma_{L}=0.15$ and 0.2) which was then subtracted according the relation
\begin{equation}
 \eta_{\rm obs}=\sqrt{\eta^2-\sigma^{2}_{L^2}}=\sqrt{\eta^2-4\sigma^{2}_{L}}
 \label{noise}
\end{equation}
if $\eta \geq2\sigma_{L}$, else $\eta_{\rm obs}=0$. For the real data, we have only an estimate of the noise level and we have to try to subtract different values from the data to find the best results. We also tested synthetic data without any noise ($\sigma_{L}=0$).

After applying our model on these data, we should obtain uniform distributions of the ecliptical longitudes $\lambda$ and latitudes $\sin\beta$. This was satisfied for the resulting distribution of $\lambda$, however, the distribution of latitudes showed preference for high $\beta$. The possible explanation is that we did not include the uncertainties following from Hapke's model and from the assumption that asteroids are triaxial ellipsoids. That means, for example, that for synthetic data without any noise and for an asteroid with $\beta=0$, there will be still some changes in brightness which our model will interpret as nonzero $\beta$. To improve the model we added a new parameter which we called model noise $\sigma_{\rm model}$. Then the equation (\ref{noise}) had to be changed to
\begin{equation}
 \eta_{\rm obs}=\sqrt{\eta^2-4\sigma^{2}_{L}-\sigma^2_{\rm model}}
 \label{noise_new}
\end{equation}
if $\eta \geq\sqrt{4\sigma^2_{L}+\sigma^2_{\rm model}}$, else $\eta_{\rm obs}=0$. 

\begin{figure}
\centering
\includegraphics[width=\hsize]{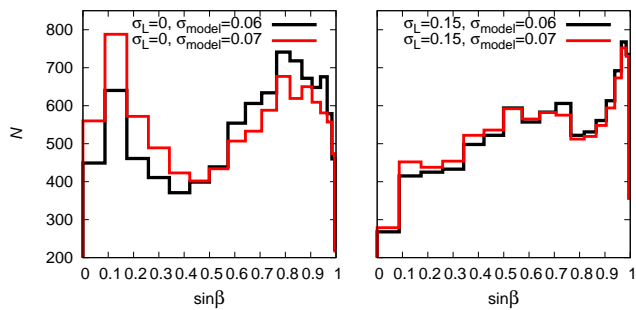}
\caption{The distributions of ecliptical latitudes $\beta$ calculated from synthetic data with noise $\sigma_L=0$ (left) and $\sigma_L=0.15$ (right) for two best values of $\sigma_{\rm model}$.}
\label{beta_synth}
\end{figure}

We tested values $\sigma_{\rm model}=0.05$, 0.06, 0.07 and 0.1. The resulting distributions of $\lambda$ were uniform independently on $\sigma_{\rm model}$. It is probably because $\lambda$ is principally determined from the mean brightness $\langle L^2\rangle$, which is comparatively more stable than the dispersion of brightness $\eta$ from which $\beta$ is determined. In Fig. \ref{beta_synth} on the left, there are shown distributions of $\sin\beta$ for two best values of $\sigma_{\rm model}$ and for the data noise $\sigma_L=0$. The distributions are clearly non-uniform, nevertheless it is the best result we obtained. When we added a noise to the synthetic data, we found, that there is no significant difference between distributions of $\sin\beta$ for $\sigma_{\rm model}=0.06$ and 0.07 (see Fig. \ref{beta_synth} on the right), therefore for the real data, we decided to use the value 0.06.

The takeaway message from our tests is that (i) determination of $\lambda$ is reliable in a statistical sense, while (ii) determination of $\beta$ is subject to systematic bias that needs to be corrected before interpreting the results.

\section{The distribution of ecliptical longitudes}

Having tested our approach and calibrated its parameters, we now construct the distribution of ecliptical longitudes for the real data from the Lowell Observatory photometric database (Bowell et al. \citeyear{2014M&PS...49...95B}). This database contains data from 11 observatories which are stored in the Minor Planet Center. The data were calibrated using the broad-band accurate photometry of the Sloan Digital Sky Survey, the accuracy is $\sim0.1-0.2$ mag. For more information about the data reduction and calibration see \citet{2011JQSRT.112.1919O}.

\begin{figure}
\centering
\includegraphics[width=\columnwidth]{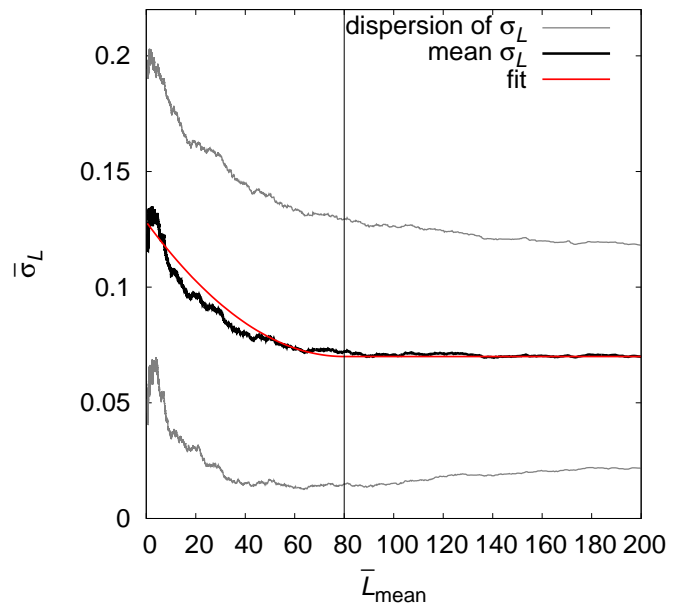}
\caption{The dependence of the mean value of noise level $\sigma_L$ on the mean value of $L_{\rm mean}$ ($L_{\rm mean}$ is mean brightness over all apparitions). Each mean value was calculated from a sample of 500 asteroids (as described in the main text). Grey lines denote dispersions of $\sigma_L$ among corresponding 500 bodies. The red line denotes the fit that was applied in the model.}
\label{noise_dependence}
\end{figure}

First, we applied our model to 765 asteroids included in DAMIT database (from the first $10\,000$ numbered asteroids, that are included in the Lowell Observatory database and satisfy the conditions on the number of apparitions and the number of measurements in one apparition) and tried different values of noise $\sigma_L$ (0.08, 0.1, 0.12, 0.15) and weight $w$ (1, 5, 25), the value of model noise was 0.06. To decide which is the best noise level and weight we compared calculated $\lambda$ and $\beta$ with $\lambda_{\rm DAMIT}$ (values from DAMIT derived with the lightcurve inversion) and $\beta_{\rm DAMIT}$, respectively. From the distributions of $\Delta\lambda=|\lambda-\lambda_{\rm DAMIT}|$ we found the best value of weight as $w=5$ and from the distributions of $\Delta\beta$ we found the best value of noise level as $\sigma_L=0.08$. However, we revealed that with this assumed data noise, the model does not produce almost any spheroidal asteroids $a/b\sim 1$. The reason is that the photometric data for less bright asteroids have higher noise level than for brighter ones. In DAMIT, there are preferentially brighter asteroids, thence the noise level 0.08 works for them, but for less bright asteroids, such noise level is underestimated. To estimate the dependence of $\sigma_L$ on $L$ we used the amplitudes $A_{\rm mag}$ of lightcurves stored in the Lightcurve Asteroid Database (LCDB)\footnote{http://www.minorplanet.info/lightcurvedatabase.html} compiled by \citet{2009Icar..202..134W}. For $A_{\rm mag}$ we can write
\begin{equation}
 A_{\rm mag}=2.5\, {\rm log}\frac{L_{\rm max}}{L_{\rm min}}=2.5\,{\rm log} \frac{L|_{\phi=0}}{L|_{\phi=\pi/2}}\,,
\end{equation}
where $L$ is given by equation (\ref{abrightness}). The normalized dispersion of brightness $\eta$, defined by equation (\ref{dispersion}), is then related with amplitude as
\begin{equation}
 \eta_{A} = \frac{1}{\sqrt{8}}\left(\frac{1}{1-A^2}-\frac{1}{2}\right)^{-1}\,,
 \label{etaa}
\end{equation}
where $A=L_{\rm min}/L_{\rm max}=10^{-0.4A_{\rm mag}}$. For 9698 asteroids included in LCDB we calculated $\eta_A$ according the eq. (\ref{etaa}) and then the appropriate noise level in data for each asteroid:
\begin{equation}
 \sigma_L = \left(\sqrt{\eta^2 - \eta_{A}^{2} - \sigma^{2}_{\rm model}}\right)/2
\end{equation}
if $\eta > \sqrt{\eta_{A}^{2} + \sigma^{2}_{\rm model}}$, else $\sigma_L=0$. To obtain the dependence of $\sigma_L$ on the mean brightness over all apparitions $L_{\rm mean}$ we calculated the running mean of $\sigma_L$ for the sample of 500 bodies. The resulting dependence, with dispersion of $\sigma_L$ among corresponding 500 bodies, is shown in Fig. \ref{noise_dependence}. We applied this dependence in our model as follows: for asteroids with $L_{\rm mean}>80$ (the brightness here is a dimensionless quantity calculated from magnitude $M$ as $L=10^{-0.4(M-15)}$) we assumed the noise level $\sigma_L=0.07$ and for asteroids less bright than 80 we calculated noise level according to the equation of parabola:
\begin{equation}
 \sigma_L=0.07+\frac{(L_{\rm mean}-80)^2}{2\times55\,000}\,.
 \label{parabola}
\end{equation}
The appropriate curve is shown in Fig. \ref{noise_dependence} (red line). We can see it does not fit the data perfectly, nevertheless, considering the dispersion of values of $\sigma_L$ (grey lines), such deviation is insignificant.

\begin{figure}
\centering
\includegraphics[width=\columnwidth]{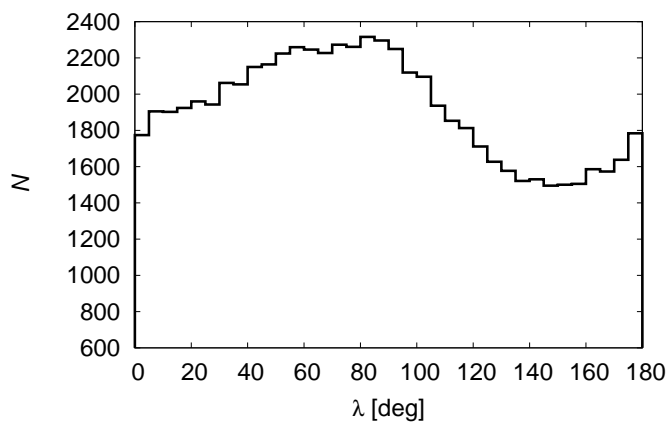}
\caption{The distribution of $\lambda$ derived for 69 053 asteroids from the Lowell Observatory photometric database with model noise $\sigma_{\rm model}=0.06$ and weight $w=5$. The values of $\lambda$ larger than $180^\circ$ map to the values $\lambda-180^\circ$, because of the symmetry of the model.}
\label{lambda_distribution}
\end{figure}

The Lowell Observatory database contains, in total, data for $326\,266$ asteroids. For $69\,053$ asteroids, there were enough apparitions and data points to calculate ecliptical longitude $\lambda$ and latitude $\beta$ (the vast majority of these asteroids belong to the first $100\,000$ numbered asteroids). For this sample, we used our model with weight $w=5$, model noise 0.06 and data noise calculated for each asteroid according to the rule described above. The resulting distribution of the ecliptical longitude $\lambda$ of asteroid pole orientation, shown in Fig. \ref{lambda_distribution}, is clearly non-uniform. As we can see, there is an excess of asteroids with $\lambda$ from $40^\circ$ to $100^\circ$ and a minimum for $\lambda \sim 150^\circ$. We calculated the Kolmogorov-Smirnov (KS) test of this distribution with a uniform one. The probability that they belong to the same parent distribution $Q_{\rm KS}$ is almost zero. A similar result was obtained by \cite{2014M&PS...49...95B}, who determined $\lambda$ from the maximum of a sinusoid curve fitting the variation of brightness.

The distribution of ecliptical latitudes $\beta$ shows strong preference for $\sin\beta\gtrsim0.9$, however, since the distribution of $\sin\beta$ for the synthetic data was not uniform (Fig. \ref{beta_synth}), the determined ecliptical latitudes are affected by biases and uncertainties which are not properly modelled here and therefore, in the following text we will study mainly the distribution of ecliptical longitudes $\lambda$.

\subsection{Searching for an explanation}

Up to now, there is no satisfactory explanation of such non-uniformity in distribution of ecliptical longitudes $\lambda$. We considered the observational and method biases described in \citet{2015P&SS..118..256M} and \citet{2015MNRAS.450..333S}, nevertheless, we found these are not able to influence our results, therefore, we were searching for some other observational biases and also geometrical and dynamical effects.

\subsubsection{Galactic plane bias}
First, we tested the influence of the measurements near galactic plane, where the stellar background is more dense and thus the measurements may have higher uncertainties. We eliminated the observations with galactic latitude $|b| < 10^\circ$ and repeated the analysis (for one asteroid there were in average about 6\% less points). The differences between computed $\lambda$ and $\lambda$ from the DAMIT database were comparable with values for the model with galactic plane, however, the non-uniformity in $\lambda$ was even larger. This result could suggest that, on the contrary, the shortage of observations near galactic plane could cause the non-uniformity of $\lambda$. However, if such a bias could influence our results, it would have been seen also in our test with synthetic data, since the geometry of observations was kept unchanged. Nevertheless, the resulting distribution of ecliptical longitudes was uniform, therefore, we believe our results are not influenced by such bias and we had to look for another explanation.

\subsubsection{Correlation with longitude of ascending node}
\begin{figure}
\centering
\includegraphics[width=7cm]{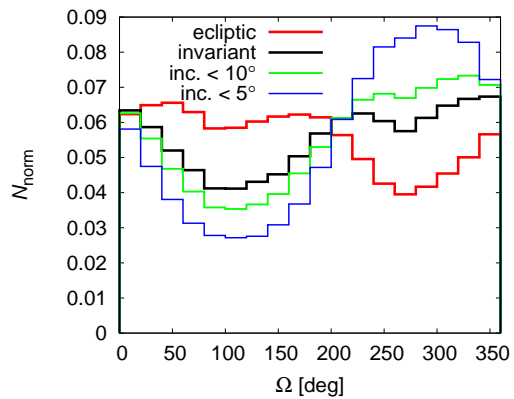}
\caption{The distribution of the longitudes of ascending node $\Omega$ (from AstOrb catalogue) for the asteroid main belt. The red curve represents the distribution in the ecliptic plane, the black curve in the invariant (Laplace) plane. The green and blue lines represent the distribution in the invariant plane for asteroids with the inclination of the orbit $I<10^\circ$ and $I<5^\circ$, respectively.}
\label{ascending_node}
\end{figure}

Next, we studied the role of the orbital longitude
      of node $\Omega$ by examining a possible correlation between
      asteroid's pole longitude $\lambda$ and $\Omega$. The orbital
      data were taken from the AstOrb catalog\footnote{ftp://ftp.lowell.edu/pub/elgb/astorb.html.}.
      Figure~\ref{ascending_node} shows distribution of $\Omega$ values for $566\,089$
      multi-opposition orbits of main-belt asteroids. 
      
      Focusing first
      on the data in the ecliptic reference system, we note that
      $\Omega$ values show over-population centered at $\simeq
      100^\circ$ value, and under-population shifted by about
      $180^\circ$, i.e. centered at $\simeq 270^\circ$ value. This
      result is not new (see, e.g., JeongAhn \& Malhotra \citeyear{2014Icar..229..236J}, and
      references therein). The reason for this non-uniformity in $\Omega$ is due to planetary perturbations. The distribution of $\Omega$ transformed to the Laplace plane shows similar non-uniformity, only shifted by $\sim180^\circ$; this is due to a slight but significant $\simeq 1.58^\circ$ tilt between the ecliptic plane and invariant plane of planets. For small-inclination orbits (i.e., whose proper inclination value is small), this effect becomes larger, as also shown in Fig.~\ref{ascending_node}. Having learned about the non-uniformity of osculating nodal
      longitudes of asteroids in the main belt we should now examine, whether
      the non-uniform distribution of their pole longitudes $\lambda$ is not a
      simple implication of the primary effect in nodes.

      First, we ran the following experiment. We divided asteroid population
      according to their value of $\Omega$ to 18 equal bins (each $20^\circ$
      wide). We found the bin that contains the smallest number $N$ of 
      asteroids, and from all other bins we randomly selected $N$ objects.
      That way, we had a sample of asteroids whose distribution of nodes
      was uniform. We examined distribution of rotation poles of this
      sub-sample, in particular the distribution of their $\lambda$ values,
      and we found it is still non-uniform, resembling that in Fig.~\ref{lambda_distribution}.
      The KS test of compatibility of the $\lambda$ distributions obtained from our sub-sample and the whole
      sample of asteroids gave us a likelihood $Q_{\rm KS}\simeq 0.90$
      that they have the same parent distribution. We repeated our
      experiment several times, creating new sub-samples, and obtained
      the same results. We also ran the same experiment in the Laplace
      reference system, but the choice of reference plane does not influence the results. These experiments suggest, non-uniformity in the
      distribution of orbital nodes does not play fundamental role in
      non-uniform distribution of pole longitudes of asteroid spins.
      
\begin{figure}
\includegraphics[width=\hsize]{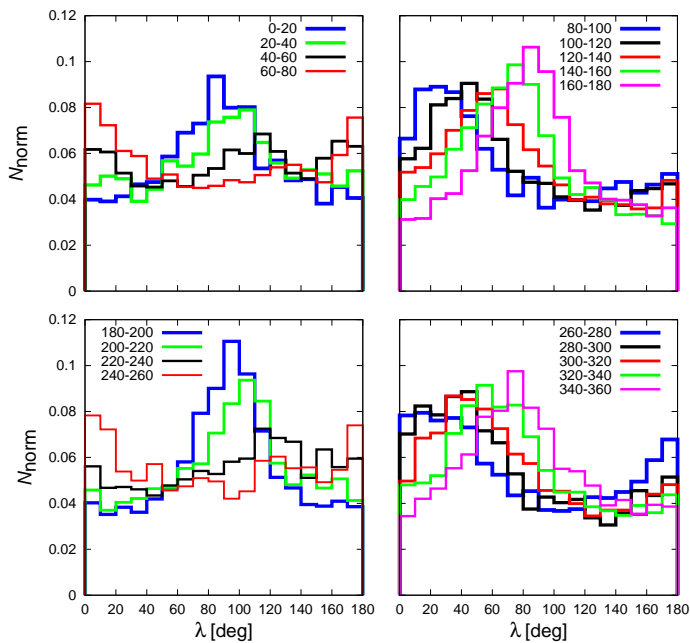}
\caption{The distributions of ecliptical longitudes $\lambda$ of poles for different longitudes of ascending node~$\Omega$.}
\label{bin_distributions}
\end{figure}
      
      Yet, we would expect some relation between $\Omega$ and $\lambda$
      should exist. For instance, plotting $\lambda$ distributions for
      asteroids in each of the $\Omega$ bins described above, we obtained
      data shown in Fig.~\ref{bin_distributions}. The results indicate that in each of the
      bins of restricted $\Omega$ values, distribution of pole longitude
      peaks at $\simeq (\Omega-90^\circ)$. This is actually understandable
      in the simplest model, in which the spin axis of each asteroid
      just uniformly precesses about the normal to its osculating orbit
      due to solar gravitational torque. We have quantitatively tested how much could a simple geometrical effect of such precession contribute to the observed dependence of ecliptical longitude $\lambda$ on node $\Omega$.
To that goal we assumed the pole position in the moving orbital plane is set with the obliquity $\epsilon$ and we chose the inclination of orbit $I$ and the longitude of ascending node $\Omega$. This initial set up was changed several times, specifically, we tested values of inclination $\sin I=0.10$, $0.15$, $0.30$, values of node $\Omega=10^\circ$, $170^\circ$, $250^\circ$ and values of obliquity $\epsilon<I$, $\epsilon>I$. Assuming a simple regular precession, we randomly picked many values of longitude $\varphi$, uniform in $0^\circ$ to $360^\circ$. We then transformed poles to the ecliptic system, determined appropriate $\lambda$ and construct a model distribution of the ecliptical longitudes. Results of these simple simulations satisfied our hypothesis of geometrical effect -- for $\epsilon<I$, the distribution of $\lambda$ was only a tight interval of values near $\simeq (\Omega-90^\circ)$, for $\epsilon>I$, the $\lambda$ values ranged the whole interval from $0^\circ$ to $360^\circ$, but with a peak at $\simeq(\Omega-90^\circ)$. However, when we, for a fixed $I$ and $\Omega$, summed distributions of $\lambda$ for values of $\epsilon$ from assumed distribution $n(\cos\epsilon)$, we reached final distribution almost uniform, far from the distributions shown in Fig. \ref{bin_distributions}. We tested $n(\cos\epsilon)$ uniform and also some unrealistic distributions, e.g. we assumed to be ten times more bodies with $\epsilon < 45^\circ$ than with $\epsilon > 45^\circ$, but with insignificant effect on the final distribution.
     
    
 Therefore, using two lines of evidence we show that the
      non-uniformity of the ecliptic $\Omega$ values together with
      only simple geometric (projection) effects cannot explain the non-uniformity in distribution of pole ecliptic
      longitudes. However, the flow
      of pole orientation in the orbit frame may be quite more
      complicated than just a simple steady precession about the
      orbital angular momentum vector. This is because of a
      possibility of resonant, spin-orbit effects described by
      Cassini dynamics (e.g., Colombo \citeyear{1966AJ.....71..891C}; Henrard \& Murigande
      \citeyear{1987CeMec..40..345H}; Vokrouhlick\'y et~al. \citeyear{2006Icar..184....1V}). In fact, the large-asteroid
      sub-group in the Koronis family, the "Slivan sample", has
      actually been identified as being captured in the
      most prominent $s_6$ Cassini resonance resulting in a common
      orientation of their pole longitudes near the stationary point
      at ecliptic longitude $\simeq 35^\circ$ (e.g., Vokrouhlick\'y
      et~al. \citeyear{2003Natur.425..147V}). Therefore, we examined whether such resonant
      effects could help us to explain the non-uniformity
      in the $\lambda$ distribution.

      However, we found the answer is negative. First, if the capture
      in the aforementioned Cassini resonance played a dominant
      role population-wise, the pole longitude distribution would be
      peaked at the stationary point of the resonance (shifted by
      some $35^\circ-40^\circ$ from the maximum seen in Fig.~\ref{lambda_distribution}).
      Next, \citet{2015A&A...579A..14V} have shown that
      (i) the capture in this resonance is generally unstable
      (especially in the inner part of the main belt), and (ii)
      its phase volume is small (few percents at maximum). The latter
      implies that expecting the spin pole located in this resonance by
      chance is very small. In order to verify these preliminary
      conclusions, we used the software described in \citet{2015A&A...579A..14V}
      to probe the expected effect. This is
      basically much more sophisticated variant of our previous
      Monte Carlo experiment where we assumed a steady precession
      in the orbit frame. Here we propagated orbit and spin evolution
      of the first $10\,000$ main belt asteroids, giving them
      random initial rotation state parameters (such as rotation
      period, pole orientation, dynamical ellipticity, etc.). We
      then numerically propagated orbit and spin evolution for tens
      of millions of years and monitored distribution of simulated
      ecliptic longitudes of the sample. We found the sample
      quickly "forgets" given initial conditions and becomes to
      fluctuate about a steady-state situation with basically
      uniform distribution of ecliptic longitudes of rotation
      poles. We repeated the numerical experiment several times with
      different initial conditions but always obtained very similar
      results.

\subsection{Distributions of $\lambda$ for groups of asteroids}

Our next step was to study the distributions of $\lambda$ for various groups of asteroids, specifically for asteroids with different sizes, different spectral types, dynamical families and asteroids in different parts of the main belt. Distributions were again compared using the KS test.

\subsubsection{Asteroids with different sizes}
\begin{figure}
\centering
\includegraphics[width=\columnwidth]{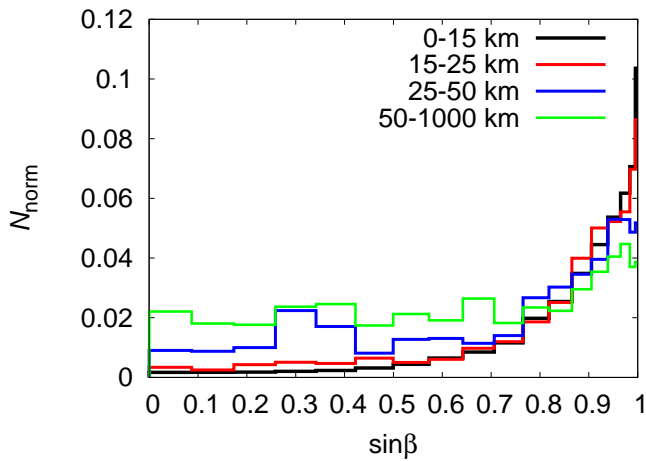}
\caption{The distribution of ecliptical latitudes $\beta$ for groups of asteroids with different sizes. All distributions are divided by the distribution for synthetic data with noise 0.15 and model noise 0.06 (Fig. \ref{beta_synth} on the right, black curve). This is the simplest de-biasing procedure we can use.}
\label{beta_real_size}
\end{figure}

We divided asteroids into eight groups according their diameters: 0--3; 3--6; 6--9; 9--12; 12--15; 15--25; 25--50; 50--1000 km (with higher diameters the number of asteroids decrease, therefore, we chose wider ranges of bins). We preferentially used diameters derived from the observations of the WISE satellite (Masiero et al., \citeyear{2011ApJ...741...68M})\footnote{http://wise2.ipac.caltech.edu/staff/bauer/NEOWISE\_pass1/.}. For asteroids not included there, we used diameters from AstOrb catalogue. We compared distributions with each other and found that the differences are not significant, which means that the data do not reveal any dependence of $\lambda$ on size. We also studied the dependence of ecliptical latitude $\beta$ on size and tried to confirm the result from \citet{2011A&A...530A.134H}. In Fig.~\ref{beta_real_size}, we can see that, despite the distributions show preference for $\sin\beta\gtrsim0.9$, with decreasing diameter $D$, there is a visible depopulation of spin axes close to the ecliptic plane, which is in agreement with findings of \citet{2011A&A...530A.134H}.

This is yet another interesting hint about the origin of the non-uniformity of $\lambda$ distribution. The affinity of latitudes toward extreme values for small asteroids clearly shows that the YORP effect has been affecting the population in a size-selective way exactly predicted by the theory. However, the distribution of the longitudes does not indicate this size-selectivity, implying the YORP effect is not the primary mechanism in the longitude story. Indeed, the theory of the YORP effect so far did not predict any significant effects for the pole longitude.

\subsubsection{Taxonomic classes}
We compared distributions of $\lambda$ between asteroids belonging to the taxonomic class C and S (using the AstOrb catalogue), which are the largest groups. The result of KS test, $Q_{\rm KS}=0.45$, indicates that there is no significant difference. The fact that the distribution of $\lambda$ is independent on the sizes and taxonomic classes simplified tests with other subpopulations.

\subsubsection{Different parts of the main belt}
\begin{figure}
\centering
\includegraphics[width=\columnwidth]{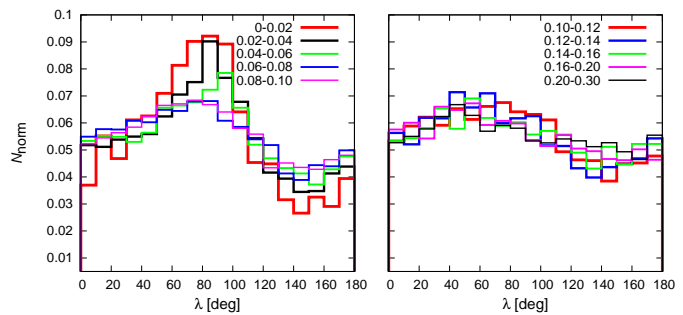}
\caption{The distributions of ecliptical longitudes for asteroids with different inclinations $\sin I$ of their orbits.}
\label{inc_parts_results}
\end{figure}

Another groups of asteroids whose distributions we studied were asteroids in different parts of the main belt. Specifically, asteroids with different inclinations $\sin I$, eccentricities $e$, and semimajor axes $a$ of their orbits. We found that the distribution of $\lambda$ is not dependent on the eccentricity, however is strongly dependent on the inclination (see Fig. \ref{inc_parts_results}). For $\sin I<0.02$ there is a huge excess of asteroids with $\lambda$ from $60^\circ$ to $100^\circ$ -- there is more than four times more bodies than for $\lambda\sim150^\circ$. With increasing $I$ the distributions are closer to the uniform distribution. Note that this result is surprising and it actually goes against the ideas about simple geometrical (projection) effects discussed in Sec. 3.1.2, suggesting that perhaps some unidentified yet dynamical effects are at play. 

We also studied the dependence of the distribution of $\lambda$ on the inclination of orbit in the invariant plane. Although the maximum of distribution for $\sin I<0.02$ is slightly lower, there is still strong dependence on the inclination.

For asteroids with $\sin I <0.04$ we constructed distributions of $\lambda$ for individual $\Omega$ bins (as in Fig. \ref{bin_distributions}). The peaks of all distribution were for $\lambda$ from $60^\circ$ to $100^\circ$, which corresponds with the distribution of $\lambda$ for small inclinations. This means, that for orbits with small inclination, the dependence of $\lambda$ on $\Omega$ is suppressed.

\begin{figure}
\centering
\includegraphics[width=8cm]{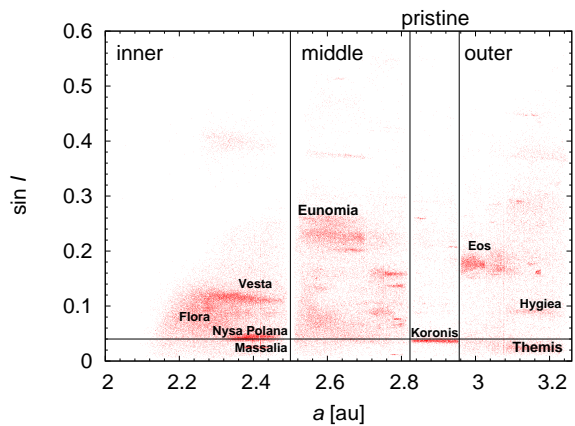}
\caption{Four parts of the main asteroid belt defined according to the proper semimajor axis $a$ (we used proper values of $a$ and $I$ from Asteroids Dynamic Site; Kne{\v z}evi{\'c} \& Milani \citeyear{2003A&A...403.1165K}). The number of object plotted here is 100 000. The horizontal line separate the region $\sin~I~<~0.04$, for which the maximum of distributions of $\lambda$ for $\lambda\sim80^\circ$ is highest. The locations of some more populous asteroid families are emphasized.}
\label{5parts}
\end{figure}

We then constructed distributions for asteroids with different semimajor axes $a$. We separated main belt into four parts (see Fig. \ref{5parts}) -- inner, middle, pristine\footnote{We adopted the word "pristine" from \citet{2013A&A...551A.117B}.}, and outer belt, which are separated by mean-motion resonances with Jupiter. To eliminate the dependence on the inclination of orbit we divided asteroids of each part into bins with different inclination (we used the same bins as in Figure \ref{inc_parts_results}) and we randomly chose such number of asteroids to have the same number of asteroids in corresponding bins of two populations. In other words, the distributions of inclination of orbit for populations, that were compared, were the same. The results of KS tests show that only the pristine zone, bracketed by the powerful mean motion resonances 5/2 and 7/3 with Jupiter at $\simeq 2.82$~au and $\simeq 2.96$~au, has significantly different distribution from the middle and outer belt ($Q_{\rm KS} < 3\times 10^{-6}$), specifically, the non-uniformity is more significant in pristine zone than in other parts. For the pair inner belt and pristine zone, the KS test gives $Q_{\rm KS} = 0.00013$.

\subsubsection{Dynamical families}
\begin{figure}
\centering
\includegraphics[width=\columnwidth]{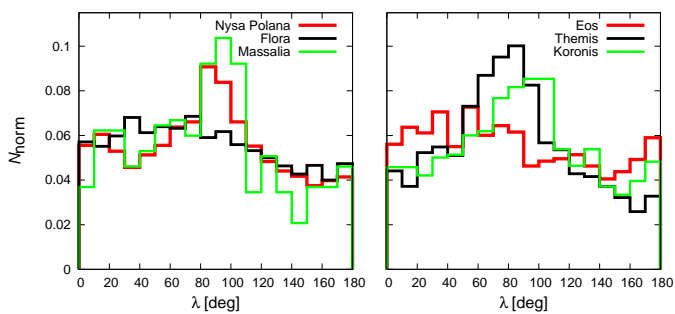}
\caption{Left: The distributions of $\lambda$ for Flora, Nysa Polana and Massalia. Right: The distributions of $\lambda$ for Eos, Themis and Koronis.}
\label{fam_outer}
\end{figure}

Finally, we studied dynamical families. The family membership of asteroids was taken from \citet{2015aste.book..297N}. Distributions of individual families were compared with the distribution of corresponding background formed with asteroids from the same part (inner, middle, pristine, outer) as the family and with inclinations of orbit from the interval defined by the members of the family. The KS test did not reveal any significant difference between any family and its background. We also compared families located approximately in the same interval of inclination (see Fig. \ref{5parts}) with each other, specifically Themis with Massalia; Vesta with Eos, Hygiea and Flora; Hygiea with Flora; Koronis with Nysa Polana. Again, the KS test showed no difference for these pairs of families. The distributions of $\lambda$ for six selected families are shown in Fig. \ref{fam_outer}. The differences we can see between the distributions are caused only due to the dependence on the inclination of orbit.

Note that the strong maximum of the $\lambda$ distribution between $\simeq (80^\circ-110^\circ)$ in the Koronis family does not fit the interval of expected librators in Slivan states described by \citet{2003Natur.425..147V} which would be shifted by about $40^\circ$ to $50^\circ$ degrees toward smaller values.

\subsection{The bootstrap method}
Formally, it is always possible to find the best ecliptical longitude $\lambda$ and latitude $\beta$ of the pole, e.g. the lowest $\chi^2$, however, the minimum can be flat and in that case $\lambda$ is not well-determined. To estimate the errors of determined longitudes we applied the bootstrap method (Davison \& Hinkley \citeyear{davison1997boostrap}) on the set of measurements for each asteroid (we used first 10 000 numbered asteroids from the Lowell Observatory database, for 9774 there were enough data points). From the set we randomly selected data to get the same number of measurements, but some of them were chosen more than once and some of them were missing. This we repeated ten times, therefore, we obtained ten modified sets of measurements and thus ten possible longitudes for each asteroid. We considered the longitude being well-determined when the maximum difference among ten values of $\lambda$ was $\leq50^\circ$. This was satisfied for 3930 from 9774 asteroids (the mean value of the largest differences for these bodies is $30^\circ$). The dependences of $\lambda$ on the longitude of ascending node $\Omega$ and on the inclination of orbit~$I$ for this new sample of 3930 asteroids did not significantly change, which means that the poorly constrained models did not cause any systematic effect to distribution of $\lambda$.

\section{Distributions of the ratio of axes $a/b$}

\begin{figure}
\centering
\includegraphics[width=6.5cm]{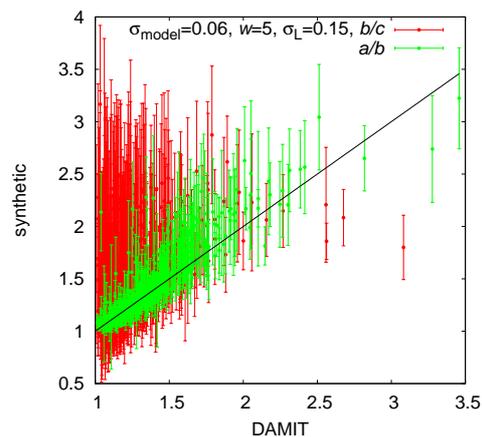}
\caption{The comparison of the values of $a/b$ and $b/c$ from DAMIT database with values calculated from synthetic data based on the models from DAMIT. One model was used more than once. The error bars are calculated as the dispersion from the mean value for one model from DAMIT.}
\label{ratios_syn}
\end{figure}

In this section, we study shapes of asteroids (specifically the ratios of axes $a/b$ and $b/c$) derived from our model. We tested our model on synthetic data as described above in Sec. 2.1, the assumed noise was 0.15. The values of ratios $a/b$ and $b/c$ obtained with our model are compared with values from DAMIT  models derived from the principal moments $I_1$, $I_2$, $I_3$ of the inertia tensor (assuming uniform density)
\begin{equation}
 \frac{a}{b}=\sqrt{\frac{I_3-I_1+I_2}{I_3+I_1-I_2}}, \frac{b}{c}=\sqrt{\frac{I_1-I_2+I_3}{I_1+I_2-I_3}}\,. 
\end{equation}
Since the values of ratios computed from our model were obtained from synthetic data based on DAMIT, they should be the same as values derived from the inertia tensor. The result is shown in Fig.~\ref{ratios_syn}. We calculated the linear (Pearson's) correlation and Spearman correlation for both ratios, the coefficients $\rho$ are summarized in Table \ref{correlation}. For the ratio $a/b$ we obtained a good correlation, the ratio $b/c$ is not so well determined.

\begin{figure}
\centering
\includegraphics[width=6.5cm]{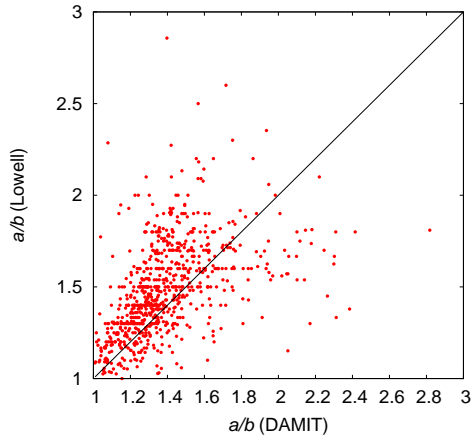}
\caption{The comparison of the values of ratio $a/b$ calculated from the real data from the Lowell database with values from DAMIT database.}
\label{ratios}
\end{figure}

For the real data, the setup was the same as described in Section 3: weight $w=5$; model noise $\sigma_{\rm model}=0.06$; and data noise $\sigma_L=0.07$, respectively for asteroids less bright than 80, $\sigma_L$ was calculated according the equation (\ref{parabola}). We compared resulting ratios $a/b$ and $b/c$ of 765 asteroids included in DAMIT with $a/b_{\rm DAMIT}$ and $b/c_{\rm DAMIT}$ and calculated correlation coefficients (see Table \ref{correlation} and also Fig. \ref{ratios}). The correlation coefficients for the ratio $b/c$ are lower than 0.1, which implies that $b/c$ is not well-determined and in following tests we will study only the ratio $a/b$. The problem to determine the ratio $b/c$ is linked with our previous result that the distribution of ecliptical latitudes $\beta$ shows preference for high values of $\beta$, especially for small bodies (see Fig. \ref{beta_real_size}), because for a spin axis with high latitude (small obliquity) we have observations only from limited range of polar aspect angles. The determination of $b/c$, however, requires observations from wide range of aspect angles.

\begin{table}
 \small
\centering
\begin{tabular}{ccc}
\hline
\hline
 & $\rho_{\rm \,linear}$ & $\rho_{\rm \,Spearman}$ \\
\hline
\rule{0cm}{2.5ex}%
synthetic data, $a/b$ & 0.88 & 0.91 \\
synthetic data, $b/c$ & 0.35 & 0.38 \\
real data, $a/b$ & 0.48 & 0.61 \\
real data, $b/c$ & 0.053 & 0.088 \\
\hline
\end{tabular}
\vspace{2mm}
\caption{\footnotesize{The linear (Pearson's) and Spearman coefficients $\rho$ of correlation.}}
\label{correlation}
\end{table}

As in Section 3.3 we used the bootstrap method to estimate errors of the ratio $a/b$. The allowed maximum difference among ten calculated values of $a/b$ was $0.25$, from 9774 asteroids remain 3819 and the mean value of the largest differences is 0.18.

\subsection{Distributions of $a/b$ for asteroids with different sizes}

\begin{figure}
\centering
\includegraphics[width=\columnwidth]{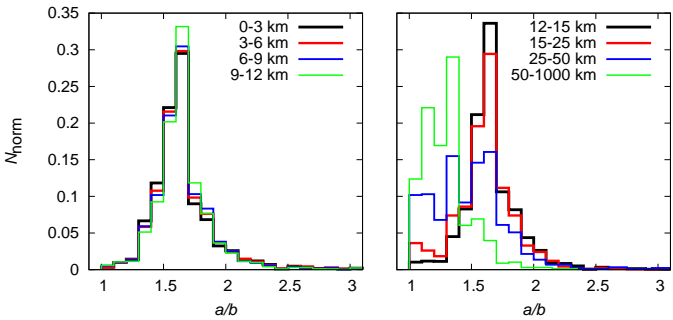}
\caption{The distributions of $a/b$ for groups of asteroids with different sizes (from 0 to 12 km on the left; from 12 to 1000 km on the right).}
\label{sizes_ab}
\end{figure}

As in the case of the ecliptic longitude $\lambda$, we studied distributions of $a/b$ for several groups of asteroids. The test for asteroids with different diameters showed, that larger asteroids ($D>25\,$km) are more spheroidal (values of $a/b$ closer to 1) and smaller asteroids are more elongated ($a/b\sim 1.6$) as is shown in Fig. \ref{sizes_ab}. The differences between distributions in Fig. \ref{sizes_ab} on the right are much bigger than the uncertainties estimated from bootstrap method 0.18. 

This dependence of $a/b$ on $D$ remained also for a smaller sample of 3570 asteroids, that were obtained from the bootstrap method as having well-determined $a/b$. Since such dependence on diameter can influence the comparison of distributions of $a/b$ of other population of asteroids we have to eliminate it in the following tests.

\citet{2016MNRAS.459.2964M} determined average axial ratio for asteroids with diameter $D<8\,$km from PanSTARRS 1 survey as 1:0.85, i.e. $a/b=1.18$, which is not in agreement with our findings. For the corresponding range of diameters, we calculated the average value of $a/b$ for asteroids in DAMIT (derived from the principal moments of the inertia tensor). The average value is $a/b_{\rm DAMIT}=1.47$, however, the sample of asteroids from DAMIT with $D<8\,$km is biased, containing preferentially more elongated asteroids, since it is easier for them to find the solution of the lightcurve inversion method. With the asteroids in DAMIT, we also checked our result, that larger asteroids are more often spheroidal -- for $D>50\,$km we obtained the average value of $a/b_{\rm DAMIT}=1.23$ (our model gives $a/b=1.29$). We can conclude that the dependence of $a/b$ on diameter is real, however our model gives higher values of $a/b$ for asteroids with $D<25\,$km. Considering the dispersion of values $\sigma_L$ in Fig. \ref{noise_dependence}, this could be because of the underestimated data noise for smaller and less bright asteroids.

\subsection{Different parts of the main belt}

Next, we studied distribution of $a/b$ for asteroids with different inclinations of their orbits. To remove the dependence of $a/b$ on diameter, we used only asteroids with $D<20\,$km. The differences between resulting distributions of $a/b$ are not so distinct as when we studied the dependence on diameter, and are comparable with the uncertainties in $a/b$.

We also compared distributions of $a/b$ for asteroids with different semimajor axes, specifically inner, middle, pristine and outer belt, using only asteroids with diameters $D<20\,$km. The differences between distributions of $a/b$ are not significant and again comparable with the uncertainties in $a/b$.

\subsection{Dynamical families and taxonomic classes}

Similarly as in Sec. 3.2.4, we compared dynamical families with their backgrounds, using again only asteroids with $D<20\,$km. We did not reveal any significant differences between distributions of $a/b$ of families and corresponding backgrounds. Also the comparison of families with each other did not show any differences larger than uncertainties in $a/b$.

\citet{2008Icar..196..135S} derived distributions of $a/b$ for eight asteroids families using data from the Sloan Digital Sky Survey (SDSS). However, their distributions are different from ours, they are often bimodal (Figures 4, 5, 6 therein) and the maximum is for $a/b\sim1.2$ (our distributions have maximum around 1.6). They also suggest a possible dependence on the age of families (old families contain more spheroidal members), but we do not observe that in our distributions. We believe, in the work they used an assumption which could influence the results. First they assumed that the rotational axis of all asteroids is perpendicular to the line of sight. Then they also tested fixed value $\beta=50^\circ$ for all asteroids.

The last populations of asteroids we compared were different taxonomic classes, specifically C and S types. The result of KS test, $Q_{\rm KS}=0.17$, did not show any difference between these two groups.

\section{Conclusions}

We developed a new method that allows to determine the orientation of
      rotational axes and equatorial axes ratio $a/b$, assuming
      a triaxial shape model, using sparse data obtained by all sky
      surveys. The goal of our approach is to provide distribution
      function of the solved-for parameters for a large sample of main
      belt asteroids rather than detailed rotational state of individual
      objects. Limitation of our method is that it provides
      (i) longitude $\lambda$ of the rotation pole in the interval
      $(0^\circ,180^\circ)$ only, with values in $(180^\circ,360^\circ)$
      transformed to $(0^\circ,180^\circ)$ by $\lambda=\lambda-180^\circ$
      rule, and (ii) absolute value of the ecliptic latitude
      $\beta$, instead of $\beta$ itself. The latter means that we cannot
      determine if the asteroid rotates in a prograde or retrograde
      sense. Additionally, our model also does not provide rotational period.
      
      We first justified our method by applying it to
      a synthetic sample of asteroids and also to a known sample of
      objects with rotational state resolved by more detailed methods
      that can be found in the DAMIT database. We observed that
      our method reproduces well distribution of the ecliptic longitudes
      and the equatorial axes ratio $a/b$ in a statistical sense. The
      uncertainties, estimated using the bootstrap method, are
      $\simeq 30^\circ$ in $\lambda$ and $\simeq 0.2$ in $a/b$ without
      systematic effects on the mean value. The determination of ecliptical 
      latitudes $\beta$ shows bias towards finding preferentially spin axes near the ecliptic pole. Our tests may,
      however, provide a rough approximation of the bias function.

      We applied then our method to $69\,053$ main belt asteroids for
      which a suitably rich and good quality set of observations were
      obtained from the Lowell Observatory database. The main results
      are as follows:

   \begin{enumerate}
      \item The distribution of $\lambda$ is non-uniform, with an excess of asteroids having $\lambda$ values between
      $60^\circ$ and $100^\circ$. Similarly, there is a deficiency
      of asteroids having $\lambda$ values between $130^\circ$ and
      $160^\circ$. Curiously, our tests revealed a correlation
      of this non-uniformity with orbital inclination: asteroids
      with very low-inclination orbits ($\sin I \leq 0.04$) show
      the effect more significantly than asteroids with higher-inclination
      orbits.
\item While not a primary result from our paper, we also determined
      distribution of the absolute value of sine of ecliptic latitude
      $|\sin\beta|$. We confirm previously reported results that
      asteroids with size $D\leq 25$~km have their pole latitude tightly
      clustered about the poles of ecliptic. This is due to the YORP
      effect that makes the pole latitude to asymptotically approach
      the extreme values.
\item We also found that small main belt asteroids ($D\leq 25$~km)
      are more elongated, with a median of ratio $a/b\simeq 1.6$, compared
      to the large asteroids ($D\geq 50$~km), which have a median of ratio
      $a/b\simeq 1.3$.
\item We also analyzed our results for populations in different asteroid
      families. As to the $\lambda$ distribution, they mainly derive from
      their inclination value of the aforementioned inclination
      dependence. For instance, the low-inclination families such as
      Massalia or Themis have the strongest non-uniformity of the
      $\lambda$ distribution in our results.
\end{enumerate}
      Using a more detailed method, we confirmed the previously reported
      unexpected non-uniformity in distribution of ecliptic longitude of
      spin axes of the main belt asteroids. We tested various hypotheses
      of its origin, but we had to reject them, proving that the proposed
      processes would not lead to a significant enough non-uniformity.
      Therefore, this result remains enigmatic and requires further
      analysis. In particular, it would
      be very useful if more detailed methods of spin state and shape
      inversion from astronomical data confirmed this result and provided
      more details. Note, for instance, that methods both in \citet{2014M&PS...49...95B} and here are not able to discriminate between the
      prograde- and retrograde-rotating asteroids. It would be important
      to see, if the excess in $\lambda$ values at about $80^\circ$
      concerns equally well both classes, or whether it is preferentially
      associated with one of them. This could hint about the underlying
      processes that cause the effect. In the same way, all methods used
      so far fold the whole range of ecliptic $\lambda$ values to
      a restricted interval $(0^\circ,180^\circ)$. This is because
      of their intrinsic drawback of not distinguishing data for
      $\lambda$ and $\lambda+180^\circ$ cases. Yet, breaking this
      uncertainty may also help to disentangle the underlying physical
      causes of the non-uniformity.

Justifications of reliability of our method, by running blind tests against synthetic populations of asteroids and limited dataset for which complete models are already available, make from it a solid tool for further studies. It would be interesting to apply it to more accurate photometric data provided by Large Synoptic Survey Telescope (LSST).
      
\begin{acknowledgements}
      H. Cibulkov\'{a} and J. \v{D}urech were supported by the grant 15-04816S of the Czech Science Foundation. We are grateful to an anonymous referee for useful and constructive comments.
\end{acknowledgements}

%
 \bibliographystyle{aa} 
 \bibliography{bib_lowell} 
%






   
  



\end{document}